# Narrow-band photonic quantum entanglement with counterpropagating domain engineering

Yi-Chen Liu[1,2], Dong-Jie Guo[1,2], Ran Yang[1], Chang-Wei Sun[1], Jia-Chen Duan[1], Zhenda Xie[1*], Yan-Xiao Gong[1*] and Shi-Ning Zhu[1]

[1] *National Laboratory of Solid State Microstructures, School of Electronic Science and Engineering, School of Physics, Collaborative Innovation Center of Advanced Microstructures, Nanjing University, Nanjing 210093, China*
[2] *These authors contributed equally to this work.*

*email: xiezhenda@nju.edu.cn, gongyanxiao@nju.edu.cn

**Photonic entanglement is a crucial resource for quantum information technologies, including quantum information processing and long-distance communication. These applications call for narrow-band entanglement sources, being compatible with the photon-electron interfaces and relax the dispersion-induced entanglement-degradation in the network mediums. However, such Giga-hertz-level bandwidth is a challenge for the photonic entanglement based on conventional spontaneous parametric down-conversion. Here we introduce a state-of-art domain-engineering technology for the counterpropagating phase matching in the polarization-entanglement generation, which results in an inherent bandwidth of 7.1 GHz at telecom wavelength. This geometry also resolves the fundamental challenge for the deterministic separation of the entangled photon pairs at frequency degeneracy, so that efficient collinear interaction can be achieved in a periodically-poled KTP waveguide and results in a spectral brightness of $3.4 \times 10^3 (GHz \cdot mW \cdot s)^{-1}$. The 155 ps base-to-base Hong-Ou-Mandel dip also confirms the source bandwidth, with and a high interference visibility of $(97.1 \pm 0.59)\%$. The entanglement is measured to violate the Bell inequality by up to 18.5 standard deviations, with Clauser–Horne–Shimony–Holt S-parameter of $2.720 \pm 0.039$. The quantum state tomography further characterizes the entanglement, with fidelity $F=(95.71 \pm 0.61)\%$. These unique features establish a cornerstone for the photonic entanglement sources development and warrant broad practical quantum information applications.**

Quantum entanglement founds the basis of the quantum information technology[1-3], and so far, the most developed entangled sources is via the optical approach because of its low decoherence and high purity. However, for practical applications the entangled photon source needs to be compatible with information processing devices, where the photon-electron interaction is normally required. One important example is the memory[4-6] for quantum information, which is not only essential for quantum computation[7,8], but also necessary for long-distance quantum communication[9-12]. It is the ultimate solution to overcome the inevitable photon loss over large communication distances and regain the channel security and data rate. The bandwidth of such photon-electron interaction is fundamentally limited by the energy level of the electrons. The recent breakthrough in the solid state quantum memories has pushed this bandwidth limit to the order of Giga-hertz[13-18], though such bandwidth is still too narrow for the conventional entangled photon sources based on spontaneous parametric down-conversion (SPDC)[19,20]. Much effort has been devoted to shrink the biphoton bandwidth, such as the passive filtering[21,22] or the cavity enhancement[23]. But it either reduces the source brightness, or adds complexity and instability of the system. On the other hand, the counter-propagating phase matching[24] geometry can inherently reduce the phase matching bandwidth without cavity interactions[25,26]. It relies on the optical microstructure manufacture and such domain engineering has been demonstrated for the mirrorless optical parametric oscillation[27] and SPDC[28,29].

Here we report the first narrowband photonic polarization-entanglement generation using counter-propagating domain engineering. The state-of-art manufacture of $\Lambda$=1.3 μm poling period in a type-II PPKTP waveguide enables 7.1 GHz biphoton bandwidth at telecom wavelength, as well as the deterministic separation of the counter-propagating signal and idler photons even at wavelength degeneracy. The bandwidth is directly measured in the spectral domain with scanning-narrow-line fitters, and also confirmed with the Hong-Ou-Mandel (HOM) interference[30]. Its high visibility of (97.1 ± 0.59)% reveals the generation of high-quality identical photon pairs. With bidirectional pump, polarization entanglement can be constructed, with Clauser–Horne–Shimony–Holt S-parameter up to $S = 2.720 \pm 0.039$. The state tomography further shows a high state fidelity of $(95.71 \pm 0.61)\%$. This entangled state fulfills the bandwidth requirement of solid state quantum memories and is thus important for the quantum information processing and communication.

In the experiment, the PPKTP waveguide is designed for backward spontaneous parametric down-conversion (BSPDC) in the type-II quasi-phase matching (QPM). As illustrated in Fig. 1a, a forward pump photon can generate a forward signal and a backward idler photon with polarizations along the *y*, *y*, and *z* axes of the KTP crystal, respectively. This BSPDC process can be phase matched with the third-order reciprocal vector of the 1.3 μm poling period in the PPKTP waveguide. We mark horizontal (H) and vertical (V) polarizations to the KTP *y* and *z* axes, respectively. With bidirectional pump at frequency degeneracy, two type-II BSPDC can happen with reversed directions, thus resulting in the generation of the following polarization-entangled state:

$$|\Psi\rangle = (|H\rangle_R |V\rangle_L + e^{i\varphi}|V\rangle_R |H\rangle_L)/\sqrt{2} \tag{1},$$

where the subscripts *R* and *L* denote the right and left propagating path, respectively.

Here the phase $\varphi$ is determined by the relative path difference between the pump at different directions.

Our experiment setup is shown in Fig. 1b. The vertically-polarized pump light is from a continuous wave Ti: Sapphire laser (SolsTis). Then a half-wave plate (HWP0), rotates its polarization, which controls the power ratio of the bidirectional pump light together with the polarization beam splitter (PBS0). The relative phase $\varphi$ can be finely tuned by the optical axis angle of the HWP3 sandwiched between two 45° quarter-wave plates (QWP3 and QWP4). The pump light is split into two beams at PBS0 with the reflected light rotated back to H polarization by HWP4. Then the two pump light beams are reflected by two dichroic mirrors (DMs) and focused into the PPKTP waveguide in two opposite directions. The above state preparation setup is integrated on a solid metal housing, where the temperature is finely controlled by Peltier elements within an accuracy of milli-Kelvin level. Therefore, the phase difference $\varphi$ in the pump loop can be stabilized (For details, see supplementary information). The DMs are designed for high transmission for BSPDC outputs at telecom wavelengths, for direct output at L and R ports. After state preparation, the output photons are measured and detected by two superconducting nanowire single photon detectors (SNSPD) with efficiency over 90% at 1550 nm. Two groups of filters including a long pass filter (Thorlabs FEL0900), a band pass filter (Semrock NIR01-1570/3-25) and a Fabry–Pérot filter (FPF) (For details, see supplementary information) are used to clean the spectrum.

We first check the phase matching of the PPKTP waveguide by the backward sum frequency generation (SFG) process. A tunable semiconductor laser (Santec TSL-710) with linewidth 100 kHz is used as the fundamental light. It is first split into two beams and coupled into the waveguide through the L and R BSPDC output ports. The L and R fundamental beams are set to H and V polarizations and the SFG light is phased-matched for right propagation. By varying the laser wavelength, we record the SFG output power with a power meter (Thorlabs S154C) as a function of pump wavelength. As shown in Fig. 1c, the SFG peaks at 776.74 nm, and fits well by the sinc square function with a full width at half maximum (FWHM) bandwidth of $\Delta\lambda=56\pm0.3$ pm. With pump wavelength fixed at 776.74 nm, we can obtain the required frequency-degenerate BSPDC that is a reverse nonlinear optical process of SFG, namely, $H_{776.74nm} \rightarrow H_{1553.48nm} + V_{1553.48nm}$.

Then we study the spectral behavior of the BSPDC. For simplicity, we focus on a single–direction-pumped BSPDC by setting HWP0 to $0°$. The BSPDC spectrum measurement is performed by scanning a home-made Fabry–Pérot cavity (FPC) with a FWHM linewidth of 7.8 pm (For details, see supplementary information). This FPC transmission is much narrower than that of BSPDC, which can be tuned across the signal or idler spectrum by varying its temperature. The BSPDC spectrum is achieved in a coincidence measurement for the best signal-to-noise ratio during this FPC scan. The measured signal and idler photon spectrums are shown in Fig. 2a, with identical FWHM fitted to be 57 pm (7.1 GHz). Their central wavelength can be finely tuned to match each other by varying the PPKTP waveguide temperature. The satellite peak strength is higher than the expectations and non-symmetric, which is due to fabrication imperfections of the sub-micron domains. So we insert a pair of 100 μm thick FPFs for spectral cleaning. As shown in Fig. 2a inset, the FWHM linewidth of each FPF is measured to be about 132 pm, which is larger than the 57 pm BSPDC bandwidth and thus does not affect its spectrum in the following measurements.

The quantum feature of a two-photon source can be presented by a high-visibility quantum interference. Here it is measured using a HOM interferometer, and the narrow BSPDC bandwidth can also be characterized from the correlation time in the interference measurement. Keeping the single–direction-pumped setup, we set the polarization of the L and R output ports to be reflected at PBS1 and PBS2, so that the BSPDC light is directed to a 50:50 fiber coupler for the HOM interference. A fiber polarization controller (PC3) is used to make the polarization of the two arms identical. The relative delay ($\Delta t$) is controlled by a motorized optical delay line in the idler photon arm. The outputs of the HOM interferometer are coupled to SNSPD for coincidence measurement. Fig. 2b shows the coincidence counts in 15 seconds as a function of the $\Delta t$, and the visibility is measured to be $90.1\pm0.91\%$, or $97.1\pm0.59\%$ after subtracting the accidentals. A triangle fit of the HOM dip shows a base-to-base width of 155 ps. This result agrees well with the 7.1 GHz BSPDC bandwidth in the spectral measurement.

Finally, by rotating HWP0 away from $0°$, the polarization entanglement can be generated. Here we fix it at $45°$ for maximum entanglement generation, and HWP3 is tuned to control the relative phase $\varphi$ in Eq. 1. As an example, we adjust HWP3 to generate the following singlet state $|\Psi^-\rangle=(|H_RV_L\rangle-|V_RH_L\rangle)/\sqrt{2}$. We first characterize the entanglement via polarization correlation measurement, where HWP2 is set to $0°, 45°, \pm 22.5°$ for the projection of R photon to H, V, and $\pm 45°$ polarization. In the case of each projection, we record the coincidence counts while changing the angle of HWP1 for the simultaneous projection of L photon. The measured interference fringes are shown in Fig. 3, and all of them fit well with a sinusoidal function, and the visibilities can be simulated to be $(96.6\pm0.45)\%, (99.5\pm0.19)\%, (97.1\pm0.53)\%$ and $(97.2\pm0.53)\%$, respectively. The maximum coincidence rate exceeds 1260 Hz in 15 seconds, which corresponds to a source brightness of $3.4\times10^3 \text{Hz}/(\text{GHz}\cdot\text{mW}\cdot\text{s})$. Then we perform the Bell test with the Clauser-Horne-Shimony-Holt (CHSH)–type inequality[31] and obtain an S value of $2.720\pm0.039$, indicating a violation of the Bell inequality by 18.5 standard deviations (For details, see supplementary information).

We further harness the entanglement using quantum state tomography[32], which gives a full characterization of the output state. Such tomography measurement requires an extra pair of quarter wave plates in L and R arms, which enable a comprehensive state projection over the following states: $|H\rangle$, $|V\rangle$, $|D\rangle=(|H\rangle+|V\rangle)/\sqrt{2}$, $|R\rangle=(|H\rangle-i|V\rangle)/\sqrt{2}$. We measure the coincidence at each combination for the L and R photons, for the reconstruction of the real and imaginary parts of the density matrix $\rho_e$. With results shown in Fig. 4, the fidelity can be calculated as $F(\rho_e,|\Psi^-\rangle)=\langle\Psi^-|\rho_e|\Psi^-\rangle=(95.71\pm0.61)\%$ [33]. Above error is estimated assuming a Poissonianly fluctuation in our data measurement. The above polarization measurements confirm high-purity polarization entanglement generation, which is achieved in narrow bandwidth without cavities for the first time.

In summary, we have demonstrated a high-quality photonic quantum entanglement by the state-of-art backward domain-engineering in a PPKTP waveguide. The BSPDC geometry enables an inherent bandwidth of 7.1 GHz as well as deterministic separation of collinear frequency-degenerate polarization entanglement. Phase-stabilized bidirectional pump can be easily achieved in balanced arm loop with only temperature control. We show an example of singlet state generation with a fidelity of (95.71±0.61)%, which can be generalized to any arbitrary polarization entangled state in our setup. In this work, our fabrication is limited by the current lithography capability to the third-order QPM for the BSPDC realization, where the spectral brightness of the photonic entanglement exceeds $3.4 \times 10^3 \, \text{Hz}/(\text{GHz} \cdot \text{mW} \cdot \text{s})$. This result is already comparable with cavity-enhanced experiments in conventional QPM geometry[34], while the setup is greatly simplified without the cavity locking optics and electronics. Further improvement on the source brightness is possible with better fabrication towards first-order QPM. This technique can also be adopted into the fast-developing Lithium Niobate thin film platform, and the tight mode confinement can further boost the conversion efficiency in a much smaller footprint for large-scale integration. Therefore, counterpropagating domain engineering is a unique and powerful tool for quantum entanglement generation, which links the photonic qubits to other key elements in the quantum information processing that requires photon-electron interaction, including quantum memories, which is important for quantum information technologies.

## Methods

**CHSH Bell inequality measurement.** To test the CHSH Bell inequality we performed four sets of measurements, each of them consisting of four joint measurements with projections onto any combination of $\pm a$ (for L photon) and $\pm b$ (for R photon), respectively, where $a \otimes b \in [\sigma_z \otimes (\sigma_x + \sigma_z), \sigma_z \otimes (\sigma_x - \sigma_z), \sigma_x \otimes (\sigma_x + \sigma_z), \sigma_x \otimes (\sigma_x - \sigma_z)]$ (the chosen sets allow violation of the CHSH Bell inequality maximally). For each set, we calculate the correlation coefficient

$$E(a,b) = \frac{C(a,b) - C(a,-b) - C(-a,b) + C(-a,-b)}{C(a,b) + C(a,-b) + C(-a,b) + C(-a,-b)},$$

where S can be calculated as:

$$S = |E(a,b) - E(a,b') + E(a',b) + E(a',b')|.$$

## Acknowledgements

This work was supported by National Key R&D Program of China (No. 2017YFA0303700), the Key R&D Program of Guangdong Province (Grant No. 2018B030329001), National Natural Science Foundation of China (Contracts No. 91321312, No. 11674169, No. 11621091, No.11474050).

**FIGURES**

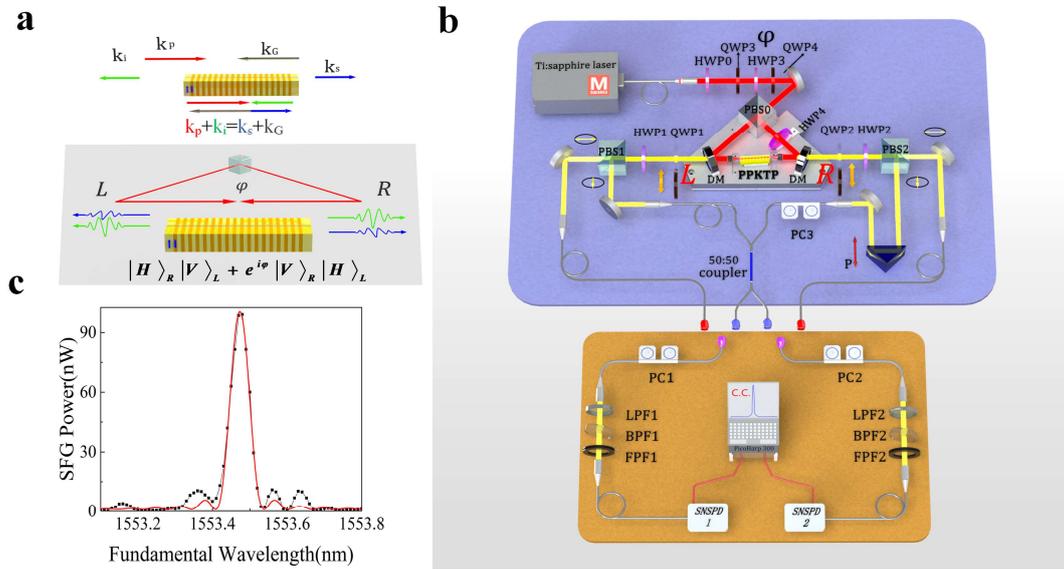

**Figure 1 |** **Scheme of the backward spontaneous parametric down-conversion and the experimental setup. a.** Geometry of the counterpropagating phase matching and entanglement generation. **b.** Experiment setup. HWP: half-wave plate; QWP: quarter-wave plate; PBS: polarization beam splitter; PC: polarization controller; LPF: long-pass filter; BPF: band-pass filter; FPF: Fabry–Pérot filer; P: prism; SNSPD: superconducting nanowire single photon detector. **c.** SFG measurement. SFG output power as a function of fundamental wavelength. The red curve is a $\mathrm{sinc}^2$-function fit.

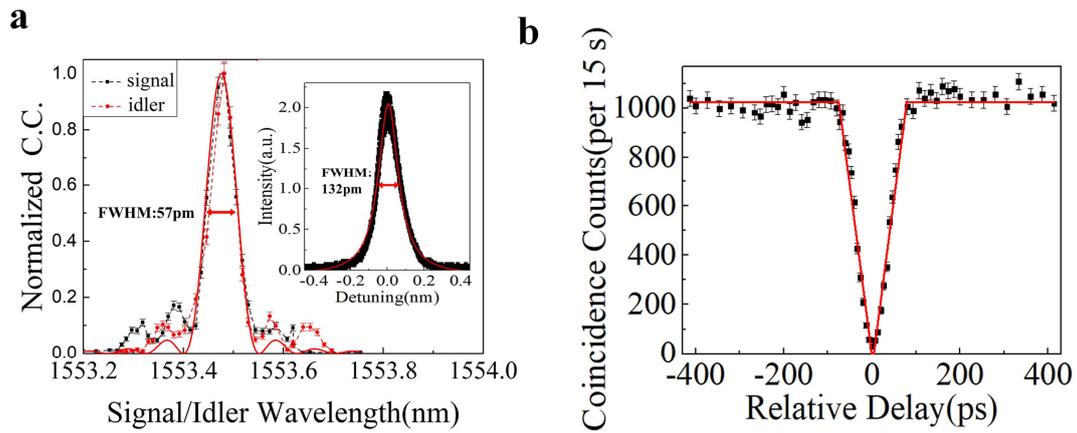

**Figure 2 |   BSPDC measurements**. **a**. Measurement of BSPDC spectrum. Black and red dots correspond signal and idler photon spectrum, respectively. The curve is fitted to sinc$^2$ functions in solid curves. Inset: Transmission spectrum of the FPF for spectral cleaning. **b**. Quantum interference measurement with HOM interferometer. The HOM dip is fitted by the red triangle for a base-to-base width of 155 ps.

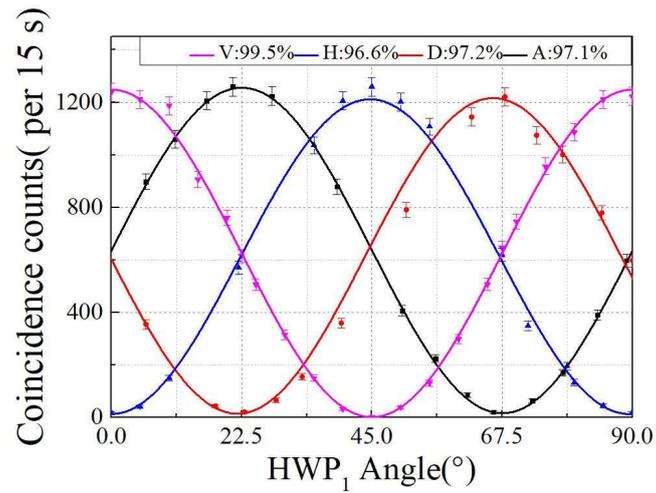

**Figure 3 | Entanglement correlation measurement.** Coincidence measurements for four different bases: H/V (blue/pink) and diagonal D/A (red/black). The results show a visibility of (96.6±0.45)%, (99.5±0.19)% in the H/V basis and (97.1±0.53)%, (97.2±0.53)% in the D/A basis.

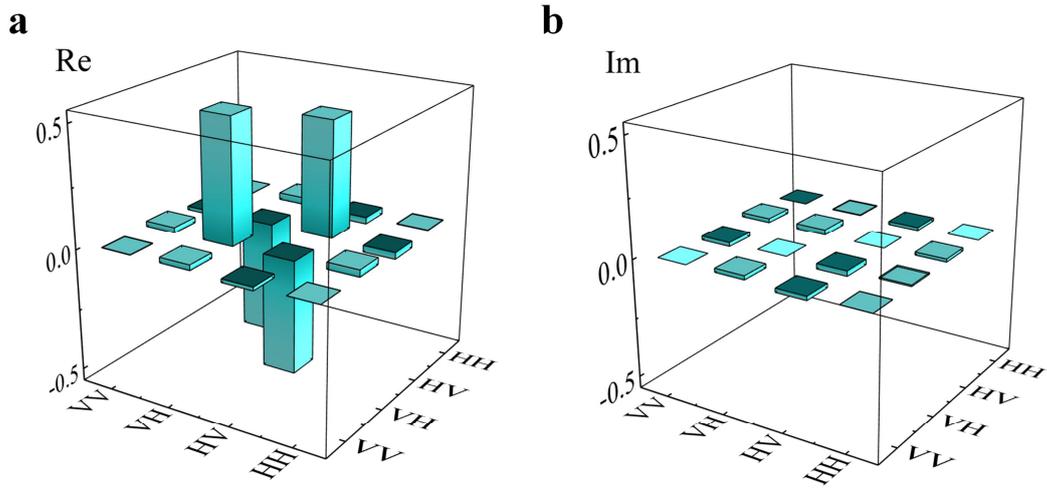

**Figure 4 |** **The real (a) and imaginary (b) parts of the reconstructed density matrix for the polarization states generated from the BSPDC process.** H and V represent the H and V bases, respectively. Without background subtractions or any corrections, the fidelity to the Bell state $|\Psi^-\rangle = (|H_R V_L\rangle - |V_R H_L\rangle)/\sqrt{2}$ is calculated as $(95.71\pm0.61)\%$.